\documentclass{article}
\usepackage[T1]{fontenc}
\usepackage{fa2020}
\usepackage{amsmath, amsfonts}
\usepackage{cite}
\usepackage{url}
\usepackage{graphicx}
\usepackage{geometry}
\usepackage{color}
\usepackage{subfig}
\usepackage[binary-units=true]{siunitx}

\usepackage{ulem}

\title{Multichannel CRNN for speaker counting: \\ An analysis of performance}

\multauthor
{Pierre-Amaury Grumiaux$^1$ \hspace{1cm} Sr\dj{}an Kiti\'c$^1$ \hspace{1cm} Laurent Girin$^2$} { \bfseries{Alexandre Guérin$^1$}\\
  $^1$ Orange Labs, France\\
$^2$ Univ. Grenoble Alpes, GIPSA-lab, Grenoble-INP, CNRS, Grenoble, France\\
{\tt\small pierreamaury.grumiaux@orange.com}
}

\begin{document}

\maketitle
\begin{abstract}
Speaker counting is the task of estimating the number of people that are simultaneously speaking in an audio recording. For several audio processing tasks such as speaker diarization, separation, localization and tracking, knowing the number of speakers at each timestep is a prerequisite, or at least it can be a strong advantage, in addition to enabling a low latency processing. In a previous work, we addressed the speaker counting problem with a multichannel convolutional recurrent neural network which produces an estimation at a short-term frame resolution. In this work, we show that, for a given frame, there is an optimal position in the input sequence for best prediction accuracy. We empirically demonstrate the link between that optimal position, the length of the input sequence and the size of the convolutional filters.
\end{abstract}
\section{Introduction}\label{sec:introduction}

Speaker counting --estimating the evolving number of speakers in an audio recording-- is a crucial stage in several audio processing tasks such as speaker diarization, localisation and tracking. It can be seen as a subtask of speaker diarization, which estimates who speaks and when in a speech segment \cite{Tranteroverviewautomaticspeaker2006,AngueraSpeakerDiarizationReview2012}. This task has been poorly addressed in the speech processing literature as a problem on its own, in a majority of the source separation and localisation methods, the number of speakers is often considered as a known and essential prerequisite \cite{VincentProbabilisticmodelingparadigms2011,MakinoBlindSpeechSeparation2007,ValinRobustLocalizationTracking2007,PerotinCRNNBasedMultipleDoA2019} or estimated by clustering separation/localisation features \cite{FallonAcousticSourceLocalization2012a,LiOnlineLocalizationTracking2019}. Speaker counting becomes even more difficult when several speech overlap, therefore it reveals particularly useful for tracking, as it can help the difficult problem of detecting the appearance and disappearance of a speaker track along time \cite{VoMultitargetTracking2015}.

In the literature of source counting, single-channel parametric methods rely on ad-hoc parameters to infer the number of speakers \cite{SayoudProposalnewconfidence2010,AraiEstimatingnumberspeakers2003}. Multichannel approaches exploit spatial information to better discriminate speakers. Classical multichannel methods are based on eigenvalue analysis of the array spatial covariance matrix \cite{WilliamsDetectionDeterminingNumber1999, NadakuditiSampleEigenvalueBased2008, YamamotoEstimationnumbersound2003}, but cannot be used in underdetermined configurations. Clustering approaches in the time-frequency (TF) domain enable to overcome this restriction \cite{balan2007estimator,ArberetRobustMethodCount2010,HiguchiUnifiedapproachaudio2015, Kounades-BastianEMalgorithmjoint2017, YangMultipleSoundSource2019}. Nevertheless, they often  turn out to be poorly robust to reverberation; moreover, they often require the maximum number of speakers as the input parameter. 

More recently, deep learning has been applied to the audio source counting problem. In \cite{WeiDeterminingNumberSpeakers2018}, a convolutional neural network is used to classify single-channel noisy audio signals into three classes : 1, 2 or 3-or-more sources. In \cite{StoterClassificationvsRegression2018}, the authors compare several neural network architectures with long short-term memory (LSTM) or convolutional layers, and also tried classification and regression paradigms. They extended their work in \cite{StoterCountNetEstimatingNumber2019} with a single-channel convolutional recurrent neural network (CRNN) predicting the maximum number of speaker occurring in a 5-second audio signal. Recently, we proposed an adaptation of this CRNN with multichannel input features to predict the number of speakers at a short-term precision on reverberant speech signals \cite{Grumiaux2020}.

In this paper, we extend our work in \cite{Grumiaux2020} by providing an empirical analysis of the speaker counting CRNN with regards to the sequence-to-one output mapping. We demonstrate that, for the best prediction on a given frame, there is an optimal choice of the decoded label within an output sequence, depending on convolutional and recurrent parameters. Past information is needed to let the LSTM converge and a few overhead frames also help for best accuracy.

\section{Speaker counting system}

The method used in this paper is the same as in \cite{Grumiaux2020}, but we shortly recall the main lines in this section.

\subsection{Input features}

To provide spatial information to the network, we use the Ambisonics representation as a multichannel input feature. The main advantages of the Ambisonics format are its ability to accurately represent the spatial properties of a soundfield,while being almost independent from the microphone type. The Ambisonics format is produced by projecting an audio signal onto a basis of spherical harmonics. For practical use, this infinite basis is truncated which defines the Ambisonics order : here, we provide first-order Ambisonics (FOA) to the network, leading to 4 channels. For a plane wave coming from azimuth $\theta$ and elevation $\phi$, and bearing a sound pressure $p$, the FOA components are given in the STFT domain by:\footnote{We use the N3D Ambisonics normalization standard \cite{DanielRepresentationchampsacoustiques2001}.}
\begin{equation}\label{eqFOA}
\begin{bmatrix} W(t,f) \\ X(t,f) \\ Y(t,f) \\ Z(t,f) \end{bmatrix} = \begin{bmatrix} 1 \\ 
\sqrt{3} \cos\theta \cos\phi \\ \sqrt{3} \sin\theta \cos\phi \\ \sqrt{3} \sin\phi \end{bmatrix} p(t,f).
\end{equation}
where $t$ and $f$ denote the STFT time and frequency bins, respectively.

The phase of $p(t,f)$ is considered a redundant information across the channels,thus we only use the magnitude of the FOA components. By stacking them, we end up with a tridimensional tensor $\mathbf{X} \in \mathbb{R}^{N_t \times F \times I}$, with $N_t$ frames, $F$ frequency bins and $I$ channels as an input feature for the neural network. We use signals sampled at $16$~kHz, a $1$,$024$-point STFT (hence $F=513$) with a sinusoidal analysis window and $50\%$ overlap. The parameter $N_t$ takes several values during our experiments, see Section~\ref{sec:analysis}.

\subsection{Outputs}

Speaker counting is considered as a classification problem with 6 classes, from 0 to 5 concurrent speakers. For the given frame, the target is encoded as a one-hot vector $\mathbf{y}$ of size 6, and the softmax function is used for the output layer. For inference, the prediction is the highest probability of the output distribution.

\subsection{Network architecture}

\begin{figure}[t]
    \centering
    \includegraphics[width=0.8\columnwidth]{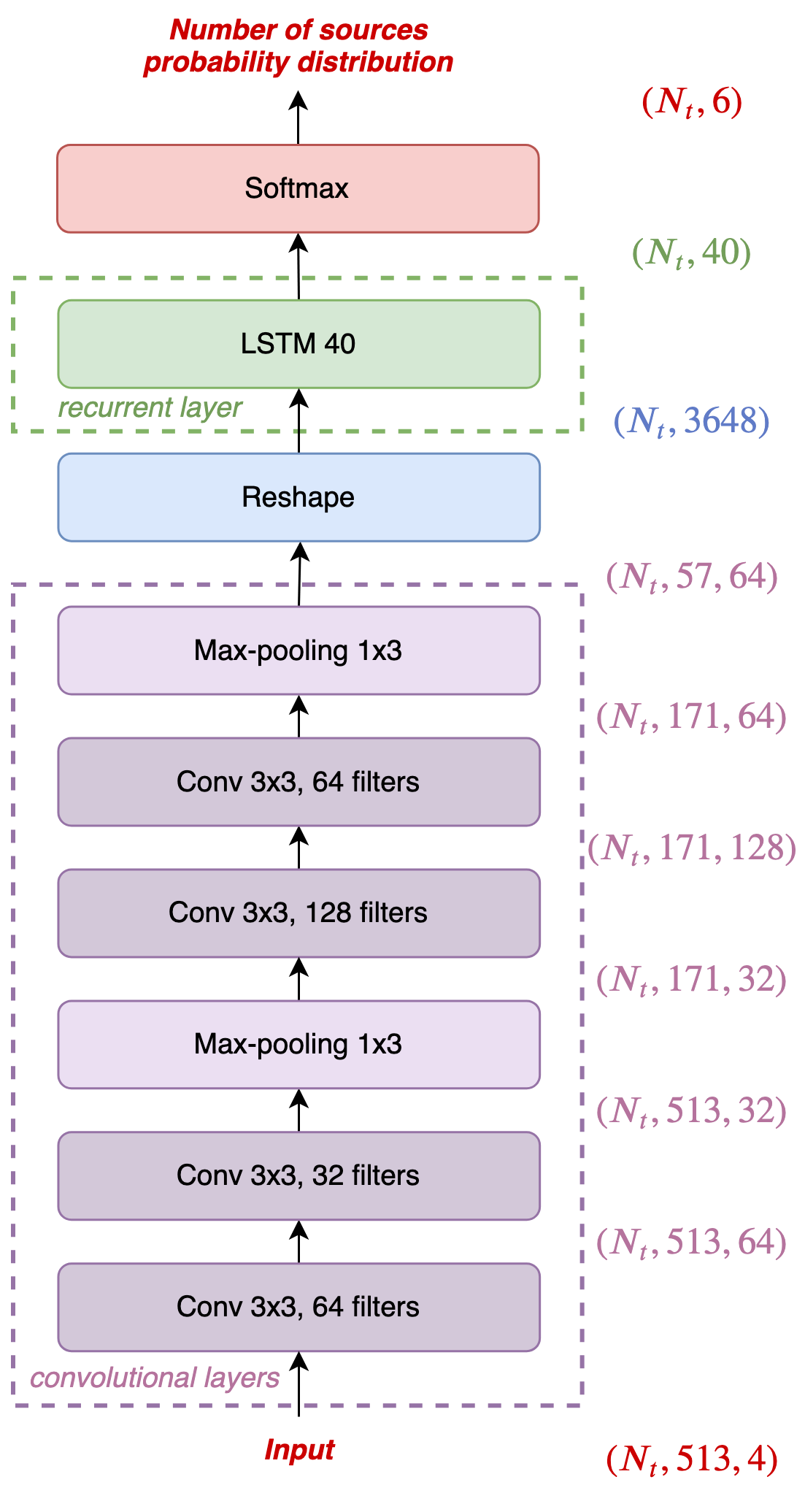}
    \caption{Architecture of the counting neural network, similar to \cite{Grumiaux2020}.}
    \label{fig:crnn}
\end{figure}

We use the same architecture as in \cite{Grumiaux2020}, illustrated in Figure~\ref{fig:crnn}. A first bloc is composed of 2 convolutional layers with 64 and 32 filters respectively, followed by a max-pooling layer, and another 2 convolutional layers with 128 and 64 filters respectively, also followed by a max-pooling layer. The filter support size $K$ (the size in both time and frequency axis) is varied as a part of the analysis in Section~\ref{sec:analysis}. 

The max-pooling operation only applies to the frequency axis to keep the temporal dimension unchanged, allowing a frame-based decision. The following layer is composed of a LSTM used in a sequence-to-sequence mapping mode (see \cite{Grumiaux2020} for more details), leading to an output of dimension $N_t \times 40$. Finally each temporal vector of dimension $40$ goes through the 6-unit softmax output layer that produces the probability distribution for each class. Therefore, this pipeline enables the network to compute a probability distribution for each frame.

\subsection{Data}

To train and test the neural network, we use synthesized speech signals comprising between 1 and 5 speakers who begin and end to talk at random times. The speech signal of each speaker is individually convolved with spatial room impulse response (SRIR) generated using the image-source method \cite{HabetsRoomimpulseresponse2006}. Then the individual wet speech signals are mixed together and a diffuse noise is added. The reader can find more details on the mixture generation algorithm in \cite{Grumiaux2020}. We end up with a total of $25$ hours of speech signals for training and $0.42$ hours for validation and test. Note that SRIR, speech and noise signals used for validation and test are never encountered during training.

\section{Performance analysis}
\label{sec:analysis}

In this section we evaluate the performance of the CRNN on the test set depending on the values of two parameters : the position $n$ of an analyzed frame within a sequence of length $N_t$ and the support size of the convolutional filters $K$.


The sequence-to-sequence nature of the LSTM layer we use provides us a way to predict the most probable class for each frame in an input sequence of $N_t$ frames. However, the amount of information available for predicting a distinct frame depends on its position $n$ within the $N_t$-frame sequence. For instance, if $n=0$ (first frame in the sequence), the prediction relies only on its content in the spectrogram plus the content of neighboring frames (because of the size of the convolutional layers), whereas for $n=N_t$ (last frame in the sequence), the prediction can fully benefit from the recurrent nature of the LSTM, by gathering information from all the previous frames in the sequence. This leads to the hypothesis that a prediction for a frame in the beginning of a sequence will be less accurate than a prediction for a frame further away in the sequence. To assess this hypothesis, we compute the accuracy of the prediction of all frames in the test set by forcing those frames to be in a same given position $n$ within the input sequence.


\subsection{Effect of frame position}

\begin{figure*}[htbp]
\centering
\subfloat[Convolutional filter size $K=3$]{\includegraphics[width=0.75\linewidth]{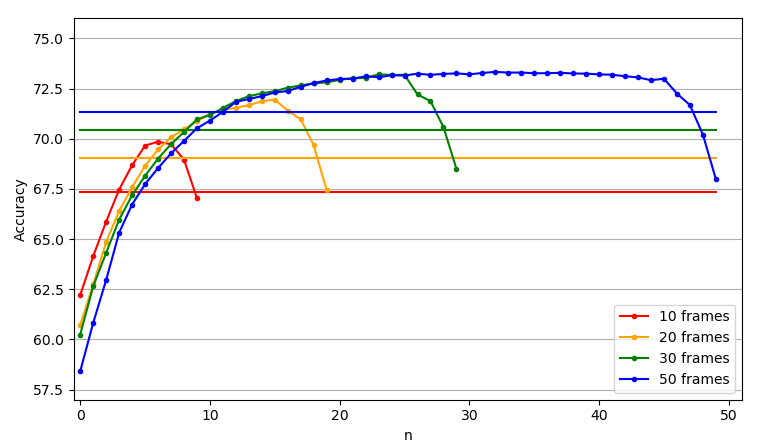}}\\
\subfloat[Convolutional filter size $K=5$]{\includegraphics[width=0.75\linewidth]{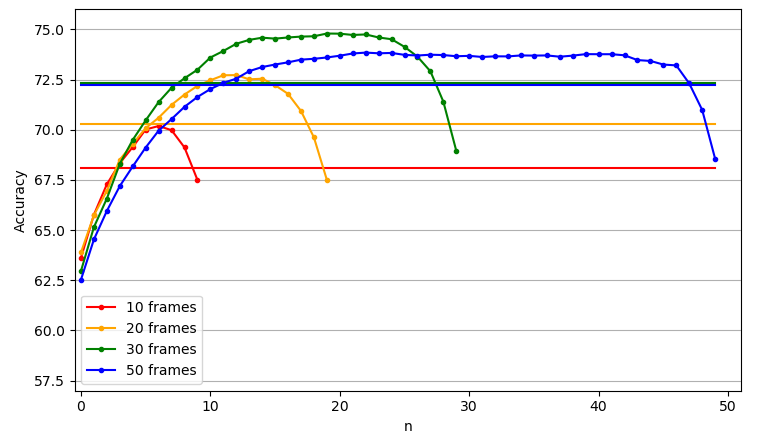}}\\
\subfloat[Convolutional filter size
$K=7$]{\includegraphics[width=0.75\linewidth]{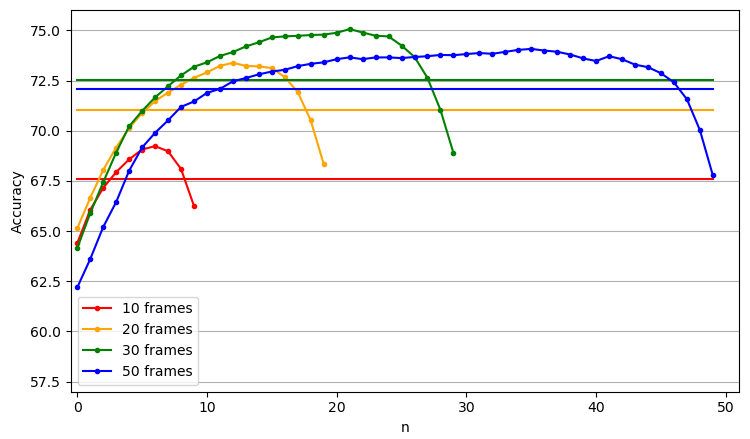}}
\caption{Average framewise speaker counting accuracy of the proposed CRNN, as a function of the position of the decoded frame in the $N_t$-sequence, for several values of $N_t$ and convolutional filter size. The horizontal bars indicated the average accuracy over the frame positions.}
\label{fig:results}
\end{figure*}

Figure~\ref{fig:results} shows the average accuracy of the CRNN on the test set, depending on the position $n$ in the sequence for the prediction of each frame and for several values of $K$. Each color corresponds to a value of $N_t$, with one curve showing the accuracy per position $n$, and one horizontal line showing the average accuracy on all positions. The results are in extent of \cite{Grumiaux2020}. We see that the CRNN is able to achieve an accuracy between $58\%$ and $75\%$, which is a good result in noisy and reverberant environment, with up to 5 speakers in the signal. As in \cite{Grumiaux2020}, we notice a trend that the average accuracy increases with the length of the sequence, but the framewise results show that the CRNN can even do better that the average performance if we avoid frames at the beginning and at the end. 

Interestingly, all curves follow a similar trend: the accuracy increases with $n$, then we observe a floor (except for $N_t=10$), then the accuracy decreases when $n$ reaches $N_t$. The first interpretation we can draw is that the LSTM layer needs a certain amount of information (several timesteps) to converge and output an optimal prediction (hence the floor). This optimal prediction is within about $69\%$ and $75\%$ accuracy, depending on the experiments.

All curves show almost the same important rise at the beginning of the sequences: For example for $K=3$ and $N_t = 30$, the accuracy goes from around $60\%$ for $n=0$ to around $69\%$ for $n=6$ which is a notable increase. At $n=6$ the accuracy begins to fall for $N_t=10$, whereas it goes on rising for other values of $N_t$. When we go further in the sequence, the increase stabilizes and reaches a floor, which is around $n=30$ for $N_t=50$. Here, the LSTM seems to have reached its optimal prediction power. So finally we can correlate the length of the input sequence to the overall accuracy with the simple reason that the LSTM needs time to converge and aggregate information to provide a better prediction. This highlights the fact that in practice we need a certain amount of past information to provide the LSTM for higher accuracy of prediction. In the following paragraph, we comment on the decrease of accuracy after the optimal floor.

\subsection{Effect of convolutional filter size}

Another interesting element in the curves is the drop of accuracy in the last values of the decoded frame position $n$, when it gets close to $N_t$. This drop of accuracy occurs for all $N_t$ values. For example for $K=3$ and $N_t=10$, accuracy gradually goes from best accuracy $69.85\%$ for $n=6$ to $67.03\%$ for last frame position $n=9$. For $N_t=30$ it goes from a maximum value of $73.19\%$ for $n=25$ to $68.50\%$ for last frame position $n=29$. In fact, for $K=3$, it seems that the peak in accuracy along the sequence appears at the frame position $n=N_t-5$ for all $N_t$ except for $N_t=10$, and for the next positions the accuracy falls.

If we do the same analysis for $K=5$, the drop seems the happen a bit earlier at position $N_t-9$, except for $N_t=10$ for which the convergence seems to condition the increase of accuracy here. For $K=7$ the phenomenon is less similar among the different values of $N_t$: for $N_t=30$ the drop begins at position $n=N_t-10$ and for $N_t=50$ it begins at position $n=N_t-16$. Yet we see that the position where the accuracy starts to drop seems to be correlated to the size of the convolutional filters. 

In fact this is due to the use of padding in the convolutional layers: For a filter size $K=3$, the spectrograms are padded with one frame of zeros when the filter is applied on an edge frame ($n=0$ or $n=N_t-1$), so that we keep the temporal dimension through all layers. This padding followed by the convolutional operation gives an edge frame with less information (since the convolution includes the void information added with padding) for the next convolutional layer, and this one will apply the same operation and so on until the $4^{th}$ and last convolutional layer. 

So after this layer the padding operation has added $1$ void frame $4$ times at the end of the spectrogram which explains the drop of accuracy $4$ frames before the end of the sequence, due to the lack of information. This can also be calculated for $K=5$: the filter size forces a padding operation of $2$ void frames for each of the $4$ convolutional layers, which leads to $2 \times 4 = 8$ void frames at the end of the spectrogram before entering the LSTM layers. We again obtain the position $n=N_t-9$ for which the accuracy begins to drop. An empirical formula can then be drawn to compute the optimal position $n_{opt}$ of the analyzed frame for best accuracy :
\begin{equation}
    n_{opt} = N_t - 2K + 1
\end{equation}
For $K=7$, the formula gives an optimal position $n=N_t-13$ before the drop in accuracy begins.

Note that this result is not perfectly observed in the curves. In particular for $N_t=10$, this padding effect is balanced by the rise of accuracy due to the LSTM convergence when accumulating information across successive frames. That is why for example, for $K= 5$ and $N_t=10$, the drop begins at $n=N_t-4$ because at that position the convergence is still in progress.

\subsection{Online vs. accuracy tradeoff}

The above analysis showed two aspects within the graphs :
\begin{itemize}
\item The LSTM needs time to converge, so that for a good speaker counting accuracy we need to provide a certain amount of past information.
\item The peak of performance is obtained for a position several frames before the end of the sequence, because after the convolutional layers the last frames suffer from padding. The number of overhead frames needed for best accuracy is $\frac{K}{2} \times 4$ where $K$ is the size of the convolutional filters. It gives the optimal position of the analyzed frame from the end of the sequence, as padding would lower accuracy in a further position.
\end{itemize}

Therefore for best speaker counting performance, a CRNN needs past information as well some overhead, depending on the recurrent and convolutional parameters.

\section{Conclusion}

In this paper we propose an analysis of the counting accuracy of a CRNN, depending on the position of the analyzed frame within the input sequence and depending on the size of the convolutional filters. We show that the LSTM indeed needs several steps to converge and provide the best possible accuracy. But although this convergence theoretically reaches its maximum at the end of the sequence, we witness a drop in accuracy towards the end of the sequence which is explained by zero-padding in the cascaded convolutional layers, which still enables us to provide a framewise prediction for source counting. The use of convolutional filters needs some overhead in the sequence. So there is a tradeoff between real-time prediction of the number of speakers and the accuracy this prediction.



\begin{thebibliography}{citations}
	
\bibitem{Tranteroverviewautomaticspeaker2006}
An Overview of Automatic Speaker Diarization Systems.
\newblock S.E. Tranter, and D.A. Reynolds.
\newblock In {\em IEEE Transactions on Audio, Speech, and Language Processing}, vol. 5, pages 1557--1565, 2006.

\bibitem{AngueraSpeakerDiarizationReview2012}
X. Anguera, S. Bozonnet, N. Evans, C. Fredouille, G. Friedland, and O. Vinyals.
\newblock Speaker diarization: A review of recent research.
\newblock In {\em IEEE Transactions on Audio, Speech, and Language Processing}, vol. 2, pages 356--370, 2012.

\bibitem{VincentProbabilisticmodelingparadigms2011}
E. Vincent, M. Jafari, S. Abdallah, M. Plumbley, and M. Davies.
\newblock Probabilistic Modeling Paradigms for Audio Source Separation.
\newblock In {\em Machine {{Audition}}: {{Principles}}, {{Algorithms}} and {{Systems}}}, pages 161--185, 2011.

\bibitem{MakinoBlindSpeechSeparation2007}
S. Makino, T. Lee, and H. Sawada.
\newblock Blind Speech Separation.
\newblock {\em Springer}, 2007.

\bibitem{ValinRobustLocalizationTracking2007}
J. Valin, F. Michaud, and J. Rouat.
\newblock Robust localization and tracking of simultaneous moving sound sources using beamforming and particle filtering.
\newblock In {\em Robotics and Autonomous System}, pages 216--228, 2007.

\bibitem{PerotinCRNNBasedMultipleDoA2019}
L. Perotin, R. Serizel, E. Vincent, and A. Gu\'erin.
\newblock {{CRNN}}-based multiple {DoA} estimation using acoustic intensity features for {Ambisonics} recordings.
\newblock In {\em IEEE Journal of Selected Topics in Signal Processing}, vol.1, pages 22--33, 2019.

\bibitem{FallonAcousticSourceLocalization2012a}
M.F. Fallon, and S.J. Godsill.
\newblock Acoustic source localization and tracking of a time-varying number of speakers.
\newblock In {\em IEEE Transactions on Audio, Speech, and Language Processing}, vol. 4, pages 1409--1415, 2012.

\bibitem{LiOnlineLocalizationTracking2019}
X. Li, Y. Ban, L. Girin, X. Alameda-Pineda, and R. Horaud.
\newblock Online localization and tracking of multiple moving speakers in reverberant environments.
\newblock In {\em IEEE Journal of Selected Topics in Signal Processing}, vol. 1, pages 88--103, 2019.

\bibitem{VoMultitargetTracking2015}
B. VO, M. Mallick, Y. Bar-shalom, S. Coraluppi, R. Osborne, and R. Mahler.
\newblock Multitarget tracking.
\newblock In {\em Wiley {{Encyclopedia}} of {{Electrical}} and {{Electronics Engineering}}}, pages 1--15, 2015.

\bibitem{SayoudProposalnewconfidence2010}
H. Sayoud, and S. Ouamour.
\newblock Proposal of a New Confidence Parameter Estimating the Number of Speakers -- {A}n Experimental Investigation.
\newblock In {\em Journal of Information Hiding and Multimedia Signal Processing}, vol. 2, pages 101--109, 2010.

\bibitem{AraiEstimatingnumberspeakers2003}
T. Arai.
\newblock Estimating the number of speakers by the modulation characteristics of speech.
\newblock In {\em IEEE International Conference on Acoustics, Speech and Signal Processing 2003.}

\bibitem{WilliamsDetectionDeterminingNumber1999}
D. Williams.
\newblock Detection: Determining the number of sources.
\newblock In {\em Digital {{Signal Processing Handbook}}}, 1999.

\bibitem{NadakuditiSampleEigenvalueBased2008}
R.R. Nadakuditi, and A. Edelman.
\newblock Sample eigenvalue based detection of high-dimensional signals in white noise using relatively few samples.
\newblock In {\em IEEE Transactions on Signal Processing}, vol.7, pages 2625--2638, 2008.

\bibitem{YamamotoEstimationnumbersound2003}
K. Yamamoto, F. Asano, W.F.G van Rooijen, E.Y.L Ling, T. Yamada, and N. Kitawaki.
\newblock Estimation of the Number of Sound Sources Using Support Vector Machines and Its Application to Sound Source Separation.
\newblock In {\em IEEE International Conference on Acoustics, Speech and Signal Processing 2003.}

\bibitem{balan2007estimator}
R. Balan.
\newblock Estimator for number of sources using minimum description length criterion for blind sparse source mixtures.
\newblock In {\em International Conference on Independent Component Analysis and Signal Separation 2007}.

\bibitem{ArberetRobustMethodCount2010}
S. Arberet, R. Gribonval, and F. Bimbot.
\newblock A robust method to count and locate audio sources in a multichannel underdetermined mixture.
\newblock In {\em IEEE Transactions on Signal Processing}, vol. 1, pages 121--133, 2010.

\bibitem{HiguchiUnifiedapproachaudio2015}
T. Higuchi, and H. Kameoka.
\newblock Unified Approach for Audio Source Separation with Multichannel Factorial {{HMM}} and {{DOA}} Mixture Model.
\newblock In {\em European {{Signal Processing Conference}} 2015.}

\bibitem{Kounades-BastianEMalgorithmjoint2017}
D. Kounades-Bastien, L. Girin, X. Alameda-Pineda, S. Gannot, and R. Horaud.
\newblock An EM Algorithm for Joint Source Separation and Diarisation of Multichannel Convolutive Speech Mixtures.
\newblock In {\em IEEE International Conference on Acoustics, Speech and Signal Processing 2017}.

\bibitem{YangMultipleSoundSource2019}
B. Yang, H. Liu, C. Pang, and X. Li.
\newblock Multiple Sound Source Counting and Localization Based on TF-Wise Spatial Spectrum Clustering.
\newblock In {\em IEEE/ACM Transactions on Audio, Speech, and Language Processing 2019.}

\bibitem{WeiDeterminingNumberSpeakers2018}
H. Wei, and N. Kehtarnavaz.
\newblock Determining the number of speakers from single microphone speech signals by multi-label convolutional neural network.
\newblock In {\em IEEE Conference of the Industrial Electronics Society 2018}.

\bibitem{StoterClassificationvsRegression2018}
F.R. St\"oter, S. Chakrabarty, B. Edler, and E.A.P. Habets.
\newblock Classification vs. regression in supervised learning for single-channel speaker count estimation.
\newblock In {\em {{IEEE International Conference}} on {{Acoustics}}, {{Speech}} and {{Signal Processing}} 2018}.

\bibitem{StoterCountNetEstimatingNumber2019}
F.R. St\"oter, S. Chakrabarty, B. Elder, and E. Habets.
\newblock {CountNet}: Estimating the number of concurrent speakers using supervised learning.
\newblock In {\em IEEE/ACM Transactions on Audio, Speech, and Language Processing 2019.}

\bibitem{Grumiaux2020}
P.A. Grumiaux, S. Kitic, L. Girin, and A. Gu\'erin.
\newblock High-Resolution Speaker Counting In Reverberant Rooms Using CRNN With Ambisonics Features.
\newblock In {\em European Signal Processing Conference 2020}.

\bibitem{DanielRepresentationchampsacoustiques2001}
J. Daniel.
\newblock Repr\'esentation de champs acoustiques, application \`a la transmission et \`a la reproduction de sc\`enes sonores complexes dans un contexte multim\'edia.
\newblock Univ. Paris VI, France, 2001.






	










\end{thebibliography}


\end{document}